\documentclass[structabstract]{aa}
\usepackage{graphicx}
\usepackage{txfonts}
\usepackage{natbib}
\usepackage{color} 
\bibpunct{(}{)}{;}{a}{}{,}

\defcitealias{NSKH07}{NSKH07}	
\defcitealias{Rea09}{RBF09}	
\defcitealias{Y10}{Y10}	
\defcitealias{HKN08}{HKN08}	
\defcitealias{HKN99}{HKN99b}	
\def\Rsolar{$R_{\odot}$}
\def\Msolar{$M_{\odot}$}

\begin{document}

\title{Single degenerate supernova type Ia progenitors}
\subtitle{Studying the influence of different mass retention efficiencies}

\author{M.C.P. Bours \inst{1,2}\fnmsep\thanks{email: m.c.p.bours@warwick.ac.uk}
	\and S. Toonen \inst{1}
	\and G. Nelemans \inst{1,3}}
\institute{Department of Astrophysics, IMAPP, Radboud University Nijmegen, PO Box 9010, 6500 GL Nijmegen, The Netherlands
	\and Department of Physics, University of Warwick, Coventry CV4 7AL, United Kingdom 
	\and Instituut voor Sterrenkunde, KU Leuven, Celestijnenstraat 200D, 3001 Leuven, Belgium\\
	}
\date{Received November 5, 2012; Accepted January 21, 2013}
\abstract{There is general agreement that supernovae Ia correspond to the thermonuclear runaway of a white dwarf that is part of a compact binary, but the details of the progenitor systems are still unknown and much debated. One of the proposed progenitor theories is the single-degenerate channel in which a white dwarf accretes from a companion, grows in mass, reaches a critical mass limit, and is then consumed after thermonuclear runaway sets in. However, there are major disagreements about the theoretical delay time distribution and the corresponding time-integrated supernova Ia rate from this channel.}
{We investigate whether the differences are due to the uncertainty in the common envelope phase and the fraction of transferred mass that is retained by the white dwarf. This so-called retention efficiency may have a strong influence on the final amount and timing of supernovae Ia.}
{Using the population synthesis code SeBa, we simulated large numbers of binaries for various assumptions on common envelopes and retention efficiencies. We compare the resulting supernova Ia rates and delay time distributions with each other and with those from the literature, including observational data.}
{For the three assumed retention efficiencies, the integrated rate varies by a factor 3-4 to even more than a factor 100, so in extreme cases, the retention efficiency strongly suppresses the single-degenerate channel. Our different assumptions for the common envelope phase change the integrated rate by a factor 2-3.  Although our results do recover the trend in the theoretical predictions from different binary population synthesis codes, they do not fully explain the large disagreement among them. }
{}
\maketitle

\section{Introduction}
Supernova Ia (SNIa) light curves are scalable to one prototype light curve due to the consistent production of a certain peak luminosity \citep[e.g.][]{Phi93}. This characteristic makes it relatively easy to estimate distances to observed SNeIa, which is why they can be used as standard candles in cosmology \citep[e.g.][]{LTCC91, Rie98, Per99}. SNeIa also strongly affect the Galactic chemical evolution through the expulsion of iron \cite[e.g.][]{Ber91}. In addition, SNeIa play an important role in astroparticle physics, because the accompanying shocks are prime accelerator sites for galactic cosmic ray particles \citep{BO78}. Although they are part of many research fields, their origin and the details of the underlying physical processes are not fully understood. 

It is generally accepted that SNIa events are caused by the thermonuclear explosions of carbon/oxygen (C/O) white dwarfs (WDs) with masses near the Chandrasekhar mass \citep{Nom82}. Of the two classical progenitor scenarios, we look at the single-degenerate (SD) channel \citep{Whe73}, in which a WD accretes from a companion. The double-degenerate (DD) channel describes the merger of two WDs \citep[e.g.][]{IT84, W84, Too12}. Other channels are amongst others accreting WDs from helium-rich, non-degenerate companions \citep[see e.g.][]{Wan09b}, and mergers between a WD and the core of an asymptotic giant branch star \citep{Kas11}. 

The progenitor theory should give a reliable description of the evolution of a SNIa progenitor binary. Beyond this, it should be able to explain and reproduce general features of the type Ia supernova class. Both the SD and DD scenarios have problems in matching observed features of the SNeIa events. We briefly mention the most important issues; however, see \citet{Liv00} and \citet{Wan12} for reviews. Regarding the DD scenario, a serious concern is whether the collapse of the remnant would lead to a supernova or to a neutron star through accretion-induced collapse \citep[see][]{Nom85, Sai85, Pie03, Yoo07, Pak10, Pak12, She12}. In the SD channel, a SNIa-like event is more easily reproduced in the simulations of the explosion process, although the explosion process needs to be fine-tuned to reproduce the observed spectra and lightcurves \citep[see e.g.][for a review]{Hil00}. 
Another issue is the long phase of supersoft X-ray emission SNIa SD progenitors should go through. It is unclear whether there are enough of these sources to account for the SNIa rate \citep[see][]{DiS10, Gil10, Hac10}. Archival data of known SNeIa have not shown this emission unambiguously, although there is possibly one case \citep[see][]{Vos08, Roe08, Nie10}. 
Finally, several predictions for the SD rate have been made by different groups using binary population synthesis (BPS) simulations \citep{YL00, Han04, Rea09, Men09, Mea10,Wan10,Y10,Cea11}, time-integrated rates, as well as the distribution of rates over time, the so-called delay time distribution (DTD). For an overview, see \citet{Nel12}. However, the results show a wide spread and do not agree with each other or with observational data. The exact origin of these differences is so far unclear. 

In this paper we try to uncover the reason for these differences, and we focus in particular on the efficiency with which the WD accretes and retains matter from its companion and on the common envelope (CE) phase. There are several prescriptions \citep[][hereafter NSKH07, RBF09 and Y10 respectively]{NSKH07, Rea09, Y10} in the literature for the efficiency of retaining matter by the WD. The retention efficiencies of \citetalias{NSKH07}, \citetalias{Rea09} and \citetalias{Y10} differ strongly. We study how the retention efficiency influences the SNIa rate and the DTD. Furthermore we investigate if the differences in the assumed retention efficiencies can explain the differences in the predicted SNIa rate from different BPS codes.  

Differences between the retention efficiencies arise from the uncertainty in the novae phase and the strength of the optically thick WD wind at high mass transfer rates \citep{KH94}. 
The optically thick wind stabilises mass transfer such that a CE phase can be avoided. \citep[][hereafter HKN99b]{HKN99} argues that the WD wind can interact with the envelope of the companion and strip some of the envelope mass from the donor, which stabilises the mass transfer further. The wind-stripping effect affects the retention efficiency moderately, but enlarges the parameter space for producing SNeIa effectively. However, for low metallicities the wind attenuates \citep[e.g.][]{Kob98, Kob09}, but a low-metallicity threshold of SNeIa in comparison with SNe type II have not been found in observations \citep{Pri08, Bad09}.
Regarding the novae phase, despite much progress in the understanding of classical novae, they are still poorly understood e.g. the mixing between the accreted envelope and WD \citep[e.g.][]{Den13}. Currently it is unclear if a cycle of novae outbursts removes more mass from the WD than the accreted envelope, effectively reducing the WD in mass \citep[][]{Pri86, PK95, Tow04, Yar05}. In particular, the results from helium novae are different, which affects the chain of nuclear burning in the SD channel for which first hydrogen-rich material is burned into helium-rich material and consequently the helium-rich matter into carbon-rich material. The uncertainty in helium accretion creates the strongest difference between the assumed retention efficiencies.

In Sect.\,\ref{sec_ch2:progs} the general evolution of a progenitor binary will be described. The CE phase and the growth in mass of the WD will be discussed in detail. To predict the distribution of SNIa rates for different assumptions we simulate the evolution for a large population of binaries using the population synthesis code SeBa, which is described in Sect.\,\ref{sec_ch2:methods}. Different methods for selecting those binaries that will evolve into a SNIa are also outlined. This includes a straightforward selection on binary parameters and an implementation in SeBa of the efficiency of mass retention on the WD. The resulting rates are presented in Sect.\,\ref{sec_ch2:results}, and compared with theoretical predictions from different groups in Sect.\,\ref{discussion}.

\section{Evolution of single degenerate SNIa progenitors} 
\label{sec_ch2:progs}
\subsection{Global evolution}
\begin{figure}[!hbtp]
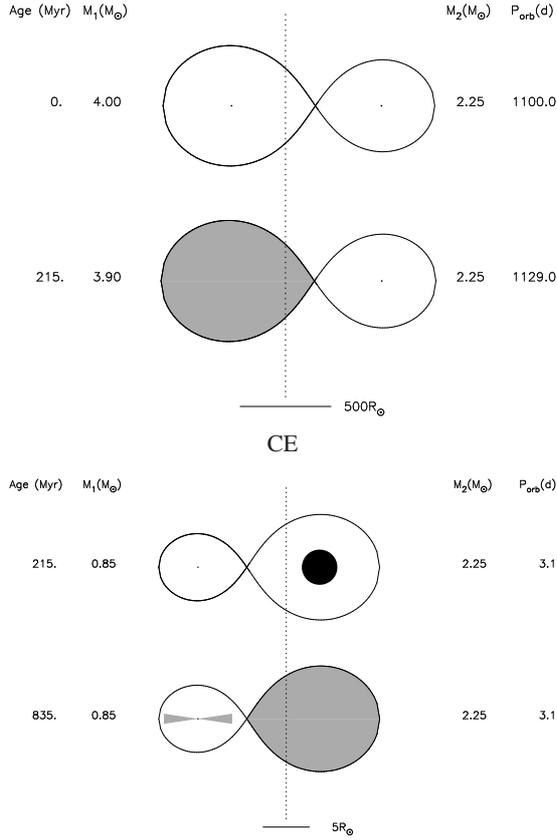

\centering
\begin{tabular}{c}
\resizebox{0.8\hsize}{!}{\includegraphics[angle = 270]{rochelobes_SD1_final.eps}} \\ 
CE \\
\resizebox{0.8\hsize}{!}{\includegraphics[angle = 270]{rochelobes_SD2_final.eps}} \\
\end{tabular}
\caption{Binary evolution for a single-degenerate SNIa progenitor. In this case the donor is a MS star. The top and bottom parts of the figure have different scales due to the common envelope phase, denoted as CE in the figure.}
\label{fig_ch2:track}
\end{figure}

The evolution of most binaries starts with two zero-age main sequence (ZAMS) stars in orbit around their common centre of mass. This is shown as the first stage in Fig.\,\ref{fig_ch2:track}, which shows the evolutionary stages a SD progenitor binary goes through. The initially more massive (hereafter primary) star has a mass of about 3-8\Msolar~and evolves into a C/O WD of about 0.6-1.2\Msolar. In order to form a compact system, binaries go through one or several mass transfer phases. The mass transfer can be stable or unstable, where the latter is described by the CE phase, the onset of which is shown in the second stage in Fig.\,\ref{fig_ch2:track}. Here the primary has evolved off the main sequence (MS) and overfills its Roche lobe. In the following CE phase the envelope of the primary star engulfs the initially less massive star, hereafter secondary. The core of the primary and secondary spiral inward through the envelope and the envelope itself is expelled. 
After the CE phase the binary consists of a degenerate C/O WD and a MS star in a close binary, see the third stage in Fig.\,\ref{fig_ch2:track}. Note that the evolutionary stages following the MS are relatively short, so that even an initially small mass difference can lead to a binary with a WD and a MS star. In spite of the importance of the CE phase for creating short period systems containing compact objects, the phenomenon is not yet well understood. In Sect.\,\ref{sec_ch2:CE} we discuss two prescriptions for the CE phase. 

If the orbital separation of the binary after the CE phase is small, the secondary will fill its Roche lobe when it evolves. Depending on the exact orbital distance this may occur when the secondary is a late MS star, a Hertzsprung gap star, or after it has evolved into a red giant.

Mass from the outer layers of the secondary will be transferred through the first Lagrangian point onto the WD. The hydrogen-rich material accumulates in a layer on the surface of the WD. When the pressure and temperature are high enough, the hydrogen layer ignites. As a result (a fraction of the) mass can be expelled from the WD and (a fraction of the) mass can be retained by the WD to form a helium layer on the WD. If the helium layer ignites, carbon and oxygen will be formed. Under the right conditions this mass too can be retained by the WD. A favourable combination of these two stages is needed for the C/O WD to be able to grow significantly in mass. In section \ref{sec_ch2:wd_growth} we will discuss in more detail the possible mechanisms of helium and hydrogen burning layers and expulsion efficiencies. As the mass of the WD approaches the critical mass limit of $M_\mathrm{SNIa}$ = 1.38\Msolar\footnote{In accordance with most of the cited papers.}, carbon ignites. Due to the degenerate nature of the WD a thermonuclear runaway takes place, leading to a SNIa event.

\subsection{Common envelope phase} 
\label{sec_ch2:CE}
The classical way to parametrise the CE phase is by the $\alpha$-formalism \citep{P76, W84} which is based on conservation of energy:
\begin{equation}
\frac{G M M_\mathrm{e}}{\lambda R} = \alpha ~ \Big{(}\frac{G M_\mathrm{c} m}{2 a_f}-\frac{G M m}{2 a_i}\Big{)} ~, 
\label{eq_ch2:alpha}
\end{equation}
where the left hand side represents the binding energy of the CE $\Delta E_\mathrm{bind}$ and the right hand side a fraction $\alpha$ of the orbital energy $\Delta E_\mathrm{orb}$ of the binary. $M$ represents the mass of the donor star and the subscripts e and c refer to the envelope and core of the donor star. The secondary mass is denoted by $m$ and is assumed not to vary during the CE phase. The $a_i$ and $a_f$ are the initial and final separation of the binary. $R$ is the radius of the donor star and $\lambda$ is a structural parameter of the envelope. All the uncertainty can be taken into one parameter, the product $\alpha\lambda$. For higher values of $\alpha\lambda$, the CE is expelled more efficiently and the change in orbital separation will be less dramatic. However, in all cases the orbital separation shrinks as a result of the CE phase. Contradicting this, observed double WD binaries seem to have mass ratios close to one \citep{M&M99}. To form a double WD pair the binary usually goes through two CE phases. One when the primary star evolves into a giant and subsequently a WD and one when the secondary evolves. In order for the mass of the WDs to be roughly equal, the conditions at the start of the first CE cannot differ much from those at the onset of the second CE. Therefore the orbital separation of these systems cannot shrink substantially during the first CE phase.

As a possible solution to this problem the $\gamma$-algorithm was introduced by \citet{Nea00}. Based on the conservation of angular momentum, angular momentum is lost in a linear way with mass according to:
\begin{equation}
\frac{\Delta J}{J} = \gamma \frac{\Delta M_\mathrm{tot}}{M_\mathrm{tot}} = \gamma \frac{M_\mathrm{e}}{M + m}, 
\label{eq_ch2:gamma}
\end{equation}
where $J$ represents the angular momentum of the binary, $M_{tot}$ is the total mass of the binary and the other parameters have the same meaning as before. Depending on the mass ratio, the orbital separation will be unchanged or decreased due to the description of the CE phase. The physical mechanism behind the $\gamma$-description remains unclear. Interesting to note here is that recently \citet{Woo12} suggested a new evolutionary model to create DWDs. These authors find that mass transfer between a red giant and a MS star can be stable and non-conservative. The effect on the orbit is a modest widening, with a result alike to the $\gamma$-description.

\subsection{Growth of the white dwarf}
\label{sec_ch2:wd_growth}
How efficiently the WD grows in mass as a result of the accretion is described by the total retention efficiency $\eta_\mathrm{tot}$. This represents the fraction of transferred hydrogen-rich matter from the companion that eventually burns into carbon-rich matter and stays on the WD. Therefore $\dot{M}_\mathrm{WD} = \eta_\mathrm{tot} |\dot{M}_\mathrm{comp}|$, where $\dot{M}_\mathrm{WD}$ represents the mass growth rate of the WD and $\dot{M}_\mathrm{comp}$ the mass transfer rate of the companion. Note that $\dot{M}_\mathrm{comp} < 0$, while $\dot{M}_\mathrm{WD} > 0$. Ultimately the retention efficiency of the transferred mass determines in which systems a SNIa event takes place and therefore shapes the SNIa rate and delay time distribution. 

\subsubsection{Hydrogen-rich accretion}
When hydrogen-rich material is transferred onto a WD, the evolution of the accreted layer depends on the mass transfer rate of the companion $\dot{M}_\mathrm{comp}$ and the WD mass $M_\mathrm{WD}$. The evolution can be split into three distinct regimes, separated by a so-called steady and critical accretion rate $\dot{M}_\mathrm{st}(M_\mathrm{WD})$ and $\dot{M}_\mathrm{cr}(M_\mathrm{WD})$ \citep{N82, HK01}. These are both functions of the WD mass and are discussed in detail below. 

For low mass transfer rates ($|\dot{M}_\mathrm{comp}| < \dot{M}_\mathrm{st}$) the hydrogen-rich matter is transferred to the WD conservatively, so that the mass transfer rate onto the WD is $\dot{M}_\mathrm{H} = |\dot{M}_\mathrm{comp}|$. The temperature in the accumulating hydrogen layer on the WD is too low to ignite the hydrogen immediately. The layer will gradually grow in mass, increasing the pressure and temperature until hydrogen ignition values are reached. Then nuclear burning will quickly spread through the layer, which starts expanding. The strength of these so-called novae \citep{Sta72} depends strongly on the amount of mass involved. This also determines how much of the processed layer is expelled from the binary and how much falls back onto the WD as helium. The accretion rate of the helium-rich matter will be denoted by $\dot M_\mathrm{He}$. We define the hydrogen retention efficiency to be the fraction of transferred mass which is burned and retained by the WD:
\begin{equation}
\eta_\mathrm{H} = \frac{\dot{M}_\mathrm{He}}{\dot{M}_\mathrm{H}}.
\end{equation}
If none of the matter stays on the WD $\eta_\mathrm{H}=0$ and if all of the transferred matter is retained by the WD as helium $\eta_\mathrm{H}=1$. In the nova regime $\eta_\mathrm{H} < 1$. For very strong novae not only the complete layer but also some of the WD itself may be expelled. In these cases $\eta_H < 0$.

Intermediate mass transfer rates are those between the steady and critical value, so that $\dot{M}_\mathrm{st} \leq |\dot{M}_\mathrm{comp}| \leq \dot{M}_\mathrm{cr}$. All of the mass lost by the companion star is transferred to the WD, so that $\dot{M}_\mathrm{H} = |\dot{M}_\mathrm{comp}|$. On the WD the hydrogen accumulates in such a way that the ignition values are reached at the bottom of the layer and the hydrogen stably burns into helium. No material is lost from the layer or the WD, therefore $\eta_\mathrm{H} = 1$ and $\dot{M}_\mathrm{He} = \dot{M}_\mathrm{H}$.

For higher mass transfer rates exceeding the critical value ($|\dot{M}_\mathrm{comp}| > \dot{M}_\mathrm{cr}$) a hydrogen red-giant-like envelope forms around the WD. If the high accretion rate is maintained a CE phase will soon follow. However, hydrogen burning on top of the WD is strong enough for a wind to develop from the WD \citep[][\citetalias{HKN99}]{KH94}. This attenuates the mass transfer rate $\dot{M}_\mathrm{H}$ and hydrogen retention efficiency $\eta_\mathrm{H}$, and ensures that the CE phase is postponed. Part of the hydrogen-rich matter, corresponding to $\dot{M}_\mathrm{cr}$, is burned into helium and stays on the WD. The remainder $\dot{M}_\mathrm{H}-\dot{M}_\mathrm{cr} = \dot{M}_\mathrm{wind}$ is carried away by the wind. In this wind regime, $\dot{M}_\mathrm{He} = \dot{M}_\mathrm{cr} = \eta_\mathrm{H} \dot{M}_\mathrm{H}$ and so $\eta_\mathrm{H} < 1$. 

A possible second effect influencing the retention efficiency in this high mass transfer rate regime, results from the wind if it reaches the companion star. There it heats up the envelope of the companion, which expands and can then be stripped off by subsequent wind \citepalias{HKN99}. It is now no longer the case that all of the mass lost by the companion is transferred to the WD. The part that is transferred to the WD is equivalent to the amount lost by the companion minus the amount stripped away by the WD wind: $\dot{M}_\mathrm{H} = |\dot{M}_\mathrm{comp}| - |\dot{M}_\mathrm{strip}|$, where $\dot{M}_\mathrm{strip} < 0$. The strength of this stripping effect is represented by the stripping parameter $c_1$, defined by
\begin{equation}
\dot{M}_\mathrm{strip} = c_1 \dot{M}_\mathrm{wind}. 
\label{eq_ch2:stripping}
\end{equation}

Details on how and when this stripping effect is taken into account are described in section\,\ref{sec_ch2:methods} and appendix\,\ref{app_ch2:retentionefficiencies}.

\subsubsection{Helium-rich accretion}
After a fraction of the hydrogen-rich matter is burned into helium-rich material, the helium-rich material accumulates in a layer on the surface of the WD. This layer sits in between the WD and the (still accumulating) hydrogen layer. For the helium-rich accretion similar regimes exist for low and medium helium accretion rates, resulting in helium novae and steady helium burning respectively. No helium wind regime exists because these high helium accretion rates cannot be reached by the hydrogen burning. The fraction of helium that is burned into carbon-rich material is represented by the helium retention efficiency $\eta_\mathrm{He}$. The total retention efficiency is then given by:
\begin{equation}
\eta_\mathrm{tot} = \eta_\mathrm{H}(\dot{M}_\mathrm{comp}) \cdot \eta_\mathrm{He}(\dot{M}_\mathrm{He}).
\label{eq_ch2:eta_tot}
\end{equation}

\section{Method} 
\label{sec_ch2:methods}
To test the viability of SD SNIa theory a population of binary stars is simulated and the corresponding theoretical SNIa rate is compared to observational results. We employ the binary population synthesis code SeBa \citep{PZV96, Nea01, Too12} for fast stellar and binary evolution computations. In SeBa stars are evolved from the ZAMS until remnant formation. At every timestep, processes as stellar winds, mass transfer, angular momentum loss, magnetic braking and gravitational radiation are taken into account with appropriate recipes. SeBa is a Monte Carlo based code in which the initial binary parameters are generated randomly according to appropriate distribution functions. The initial mass of the primary stars is drawn from a Kroupa Initial Mass Function \citep{Kroupa} which ranges from 0.1-100\Msolar~and the initial mass of the secondary from a flat mass ratio distribution with values between 0 and 1. We assume that a complete stellar population consists of 50\% binary stars and 50\% single stars. The semi-major axis of the binary is drawn from a power law distribution with an exponent of -1 \citep{Abt83}, ranging from 0 to $10^6$\Rsolar~and the eccentricity from a thermal distribution, ranging from 0 to 1 \citep{Heggie75}. For the metallicity solar values are assumed.

In this research we differentiate between four ways to determine which binaries will give rise to a SNIa event. For the first method (model NSKH07) we adopt the retention efficiencies given by \citetalias{NSKH07}, including updates from \citet[][hereafter HKN08]{HKN08}. Using SeBa in combination with these retention efficiencies, a population of binaries is simulated and those systems in which a C/O WD reaches the critical mass limit are selected as SNIa progenitors. For the second approach (model RBF09) we utilise the retention efficiencies used in \citetalias{Rea09} and for the third approach (model Y10) the efficiencies based on \citetalias{Y10}. The assumed retention efficiencies differ strongly. For the final method (model Islands) the phase of stable mass transfer onto the WD is not modelled by SeBa. Instead we select those systems that at the formation of the C/O WD lie in specific regions of the parameter space of orbital period, WD mass and companion mass ($P_\mathrm{orb}$ - $M_\mathrm{WD}$ - $M_\mathrm{comp}$). Systems in these specific regions, hereafter islands, are determined to lead to SNeIa \citepalias[][and references therein]{HKN08, HKN99}. 

In order to compare our simulated rates to the results of research groups that use the same retention efficiencies, the prescription of the CE is varied to match different BPS codes. We distinguish between two models. The first model is based on the $\alpha$-prescription assuming a value of $\alpha = 1$ and $\lambda = 0.5$ as in accordance with \citetalias{Rea09}. The second model assumes the $\gamma$-formalism with a value of $\gamma = 1.75$ \citep[see][]{Nea01}.\footnote{The $\gamma$-formalism is applied, unless the binary contains a compact object or the CE is triggered by a tidal instability \citep{Dar1879} for which the $\alpha$-formalism is used.}

\subsection{Retention efficiencies} 
\label{sec_ch2:retention_efficiencies}
\begin{figure}[!tbhp]
\resizebox{\hsize}{!}{\includegraphics{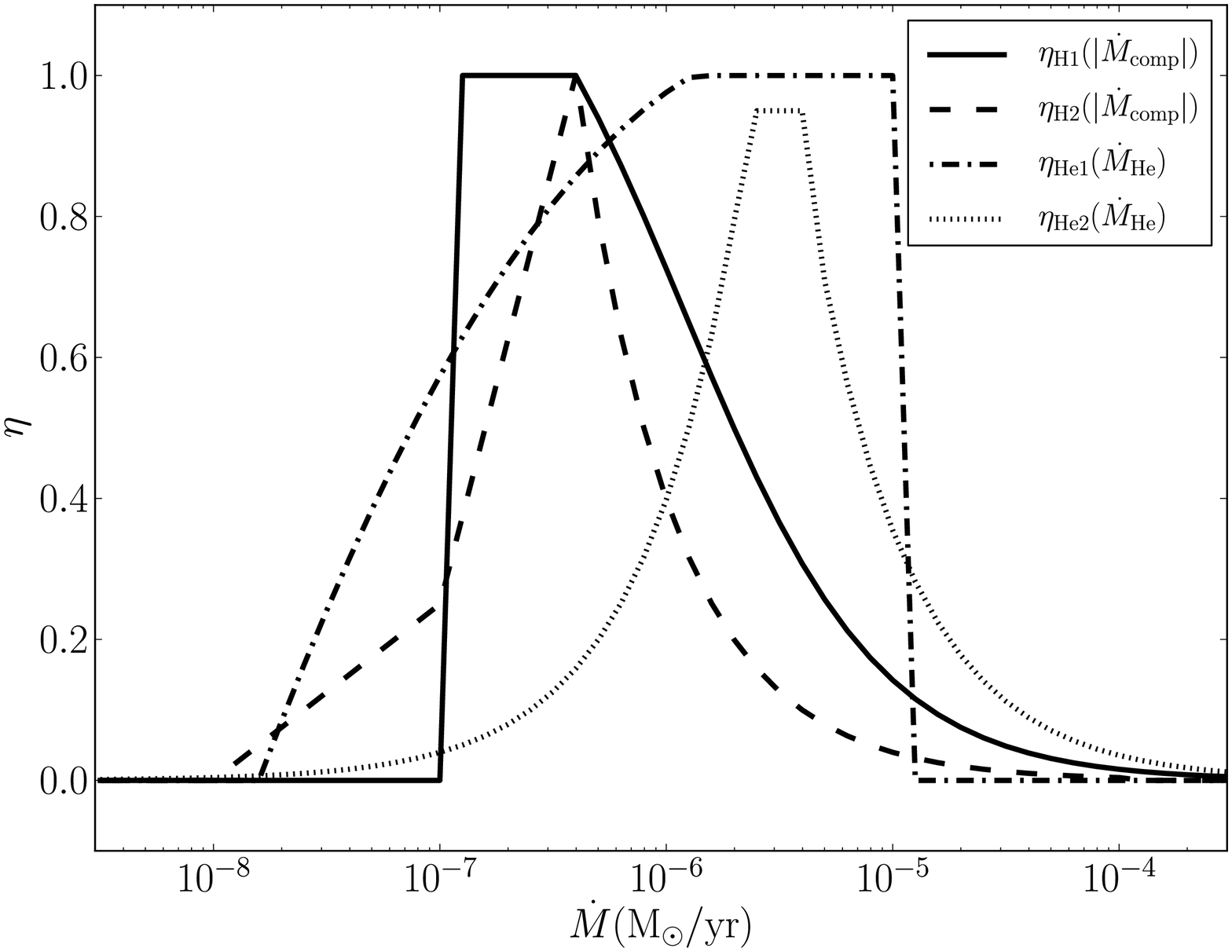}}
\caption{Examples of hydrogen and helium retention efficiencies as a function of the mass transfer rate for a WD of 1\Msolar. The prescriptions for the hydrogen retention efficiency $\eta_\mathrm{H1}$ \citepalias[solid line - ][]{NSKH07, HKN08} \& $\eta_\mathrm{H2}$ \citepalias[dashed line - ][]{Rea09, Y10} and the helium retention efficiency $\eta_\mathrm{He1}$ \citep[dot-dashed line - ][]{Hea99, KH99} \& $\eta_\mathrm{He2}$ \citep[dotted line - ][]{IT96} are shown in Appendix\,\ref{app_ch2:retentionefficiencies}. Note that the position of the peak of $\eta_\mathrm{H2}$ is dependent on the exact prescription for $\dot{M}_\mathrm{cr}$. The mass transfer rate for hydrogen accretion is given by $|\dot{M}_\mathrm{comp}|$. For helium accretion this is $\dot M_\mathrm{He}$. }
\label{fig_ch2:reteffs_HHe}
\end{figure}

In this section we describe the three prescriptions for the hydrogen and helium retention efficiencies as given by \citetalias{NSKH07} \citetalias{Rea09} and \citetalias{Y10}. The total retention efficiency is a combination of the hydrogen and helium one as described by eq.\,\ref{eq_ch2:eta_tot}. 
The retention efficiency used in \citetalias{NSKH07}, including updates from \citetalias{HKN08}, is a combination of the hydrogen retention efficiency $\eta_\mathrm{H1}$ and the helium retention efficiency $\eta_\mathrm{He1}$, see Fig.\,\ref{fig_ch2:reteffs_HHe}. For $\eta_\mathrm{H1}$ the region of steady burning occurs at mass transfer rates between
\begin{equation}
\dot{M}_\mathrm{st} = 3.1 \cdot 10^{-7} (M_\mathrm{WD} - 0.54) ~ \mathrm{M}_{\odot} \mathrm{yr}^{-1} \mathrm{~ and} \\ 
\label{eq_ch2:MstN} 
\end{equation}
\begin{equation}
\dot{M}_\mathrm{cr} = 7.5 \cdot 10^{-7} (M_\mathrm{WD} - 0.40) ~ \mathrm{M}_{\odot} \mathrm{yr}^{-1}, 
\label{eq_ch2:McritNR}
\end{equation}
which are the updated formulae from \citetalias{HKN08} with $M_\mathrm{WD}$ in units of \Msolar. This region is visible in Fig.\,\ref{fig_ch2:reteffs_HHe} as the plateau at intermediate mass transfer rates. In the wind regime 
$|\dot{M}_\mathrm{comp}| > \dot{M}_\mathrm{cr}$, the wind-stripping effect is included with $c_1=3$ (see eq.\,\ref{eq_ch2:stripping}). For the prescription of $\eta_\mathrm{He1}$ see Appendix\,\ref{app_ch2:retentionefficiencies} \citepalias[\citealp{KH99};][]{HKN99}.

The retention efficiency based on \citetalias{Rea09} combines  $\eta_\mathrm{He1}$ with $\eta_\mathrm{H2}$, see Fig.\,\ref{fig_ch2:reteffs_HHe}. In this model there is no region of steady hydrogen burning. Instead the nova regime immediately borders the wind regime. The critical mass transfer rate $\dot{M}_\mathrm{cr}$ is given by eq.\,\ref{eq_ch2:McritNR}. The stripping effect is not taken into account, hence $c_1=0$ (see eq.\,\ref{eq_ch2:stripping}). The retention efficiency in the nova regime ($|\dot{M}_\mathrm{comp}| < \dot{M}_\mathrm{cr}$) is based on an interpolation of results from \citet{PK95}. Our own fit to these results is included in Appendix\,\ref{app_ch2:retentionefficiencies}. For lower (higher) $M_\mathrm{WD}$ than assumed in Fig.\,\ref{fig_ch2:reteffs_HHe}, the retention efficiency in the nova regime stays the same between $10^{-8} < |\dot{M}_\mathrm{comp}| < 10^{-7}$ and shifts with the peak in the regime $10^{-7} < |\dot{M}_\mathrm{comp}| < \dot{M}_\mathrm{cr}$. 

\begin{figure}[!tbhp]
\resizebox{\hsize}{!}{\includegraphics{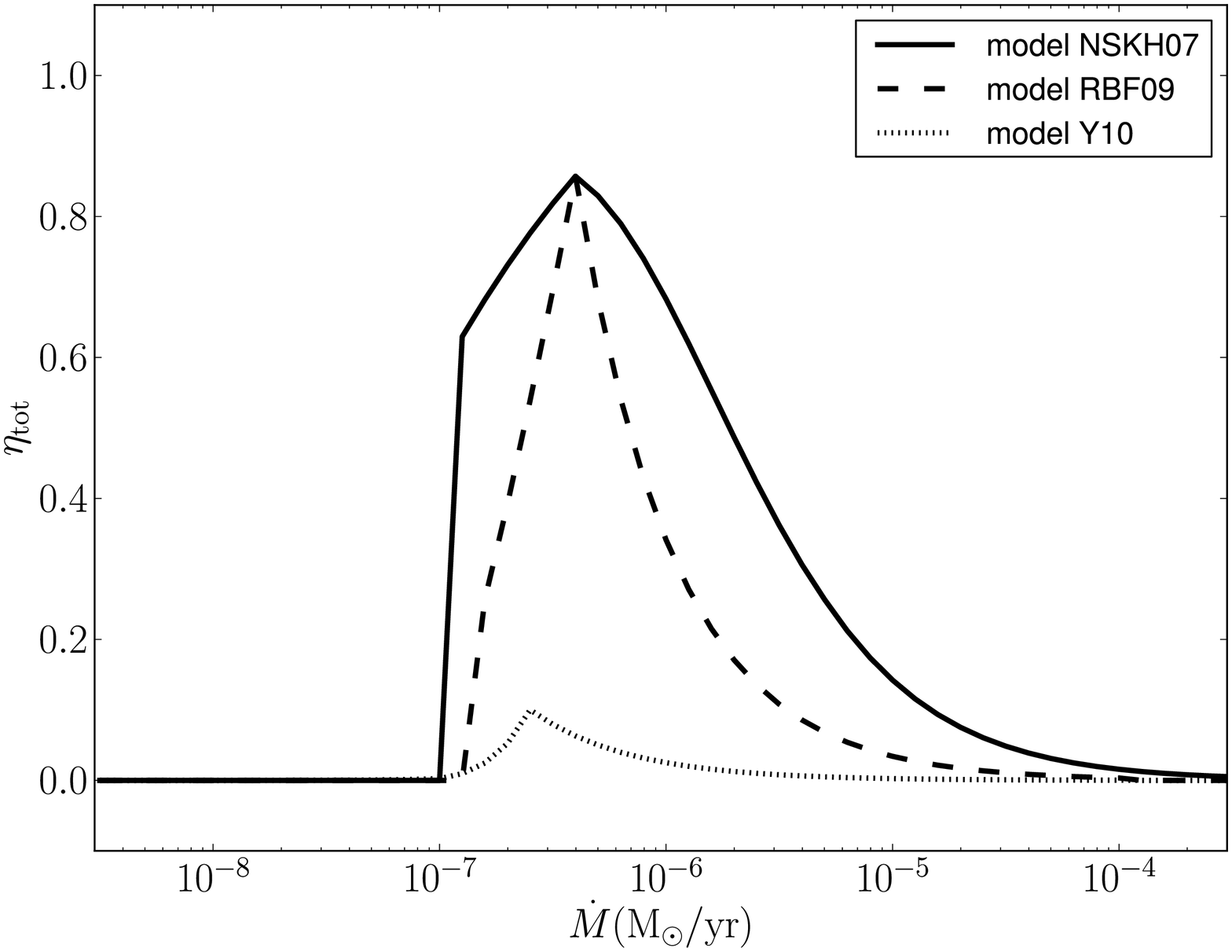}}
\caption{Total retention efficiencies, resulting from different combinations of the hydrogen and helium retention efficiencies. In this figure we have assumed $M_\mathrm{WD} = 1$\Msolar~as an example. The mass transfer rate $\dot{M} = |\dot{M}_\mathrm{comp}|$. For lower (higher) $M_\mathrm{WD}$ the maximum retention efficiency shift to lower (higher) $|\dot{M}_\mathrm{comp}|$. For a more detailed explanation of the retention efficiencies see Sect.\,\ref{sec_ch2:retention_efficiencies}. }
\label{fig_ch2:reteffs_tot}
\end{figure}

The retention efficiency based on \citetalias{Y10} combines $\eta_\mathrm{H2}$ with $\eta_\mathrm{He2}$, see Fig.\,\ref{fig_ch2:reteffs_HHe}. The critical rate for the hydrogen accretion is given by
\begin{equation}
\dot{M}_\mathrm{cr} = 10^{-9.31+4.12M_\mathrm{WD}-1.42M^2_\mathrm{WD}} ~ \mathrm{M}_{\odot} \mathrm{yr}^{-1},
\label{eq_ch2:McritY}
\end{equation}
with $M_\mathrm{WD}$ in units of \Msolar. For the retention efficiency in the hydrogen nova regime the same interpolation from \citet{PK95} is used as for the efficiency of \citetalias{Rea09}, see Appendix\,\ref{app_ch2:retentionefficiencies}. Again there is no region of steady hydrogen burning, the wind is taken into account but the stripping effect is not. Due to the different prescription for $\dot{M}_\mathrm{cr}$ the position of the peak is slightly different from the example curve.
Using eq.\,\ref{eq_ch2:McritY} as the border to the wind regime for hydrogen accretion, the peak occurs at smaller values of $|\dot{M}_\mathrm{comp}|$ for a given $M_\mathrm{WD}$, as compared to eq.\,\ref{eq_ch2:McritNR}. This reduces the total retention efficiency $\eta_\mathrm{tot}$ substantially, as we will see in Fig.\,\ref{fig_ch2:reteffs_tot}. Following \citetalias{Y10}, the prescription for $\eta_\mathrm{He2}$ was taken from \citet{IT96}.

The exact prescriptions of the hydrogen and helium retention efficiencies for \citetalias{NSKH07}, \citetalias{Rea09} and \citetalias{Y10} are detailed in Appendix\,\ref{app_ch2:retentionefficiencies}. Fig.\,\ref{fig_ch2:reteffs_tot} shows the large variety in the total retention efficiency for the three prescriptions as a function of mass transfer rate of the companion. The most optimistic retention efficiency is that of \citetalias{NSKH07} and the most pessimistic that of \citetalias{Y10}. 
Note that in the regime with strong novae the retention efficiency $\eta_\mathrm{tot} \leq 0$. Here we have set $\eta_\mathrm{tot} = 0$ in this regime for simplicity as these systems are not part of the SD channel.

A complication of this method is that the instantaneous mass transfer rates in binary population synthesis codes, such as SeBa, is only an approximation to the one derived from detailed stellar evolution codes. Therefore, we also implemented a hybrid method in which progenitors are selected from binary population synthesis according to results from the literature based on more detailed mass transfer tracks. These tracks can be calculated by an analytical approach e.g. \citetalias{HKN99} or by a detailed binary evolution code e.g. \citet{Li02}, \citet{Han04} and \citetalias{HKN08}.

\subsection{Islands}
\begin{figure}[!hbtp]
\centering
\includegraphics[width=0.24\textwidth]{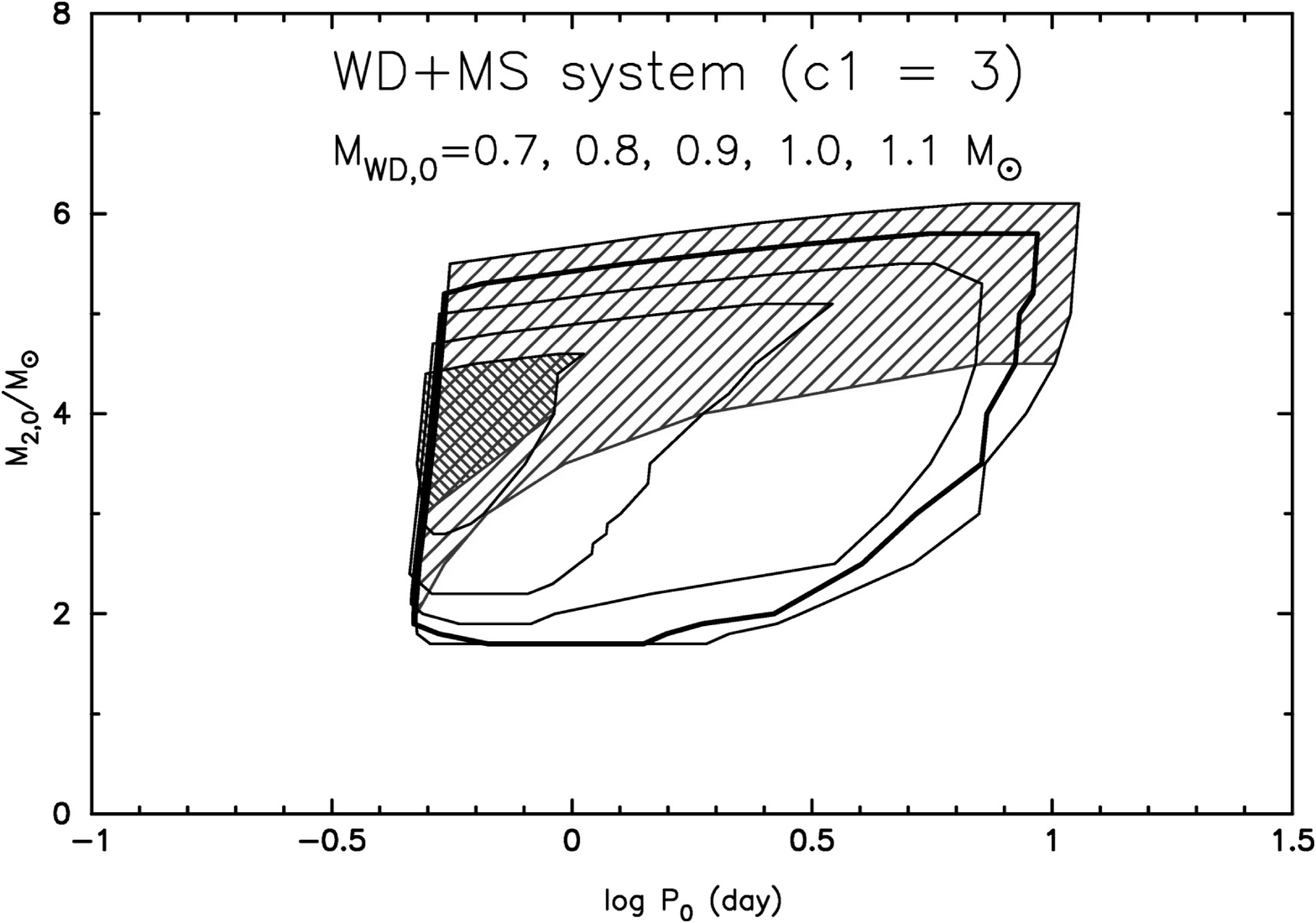}
\includegraphics[width=0.24\textwidth]{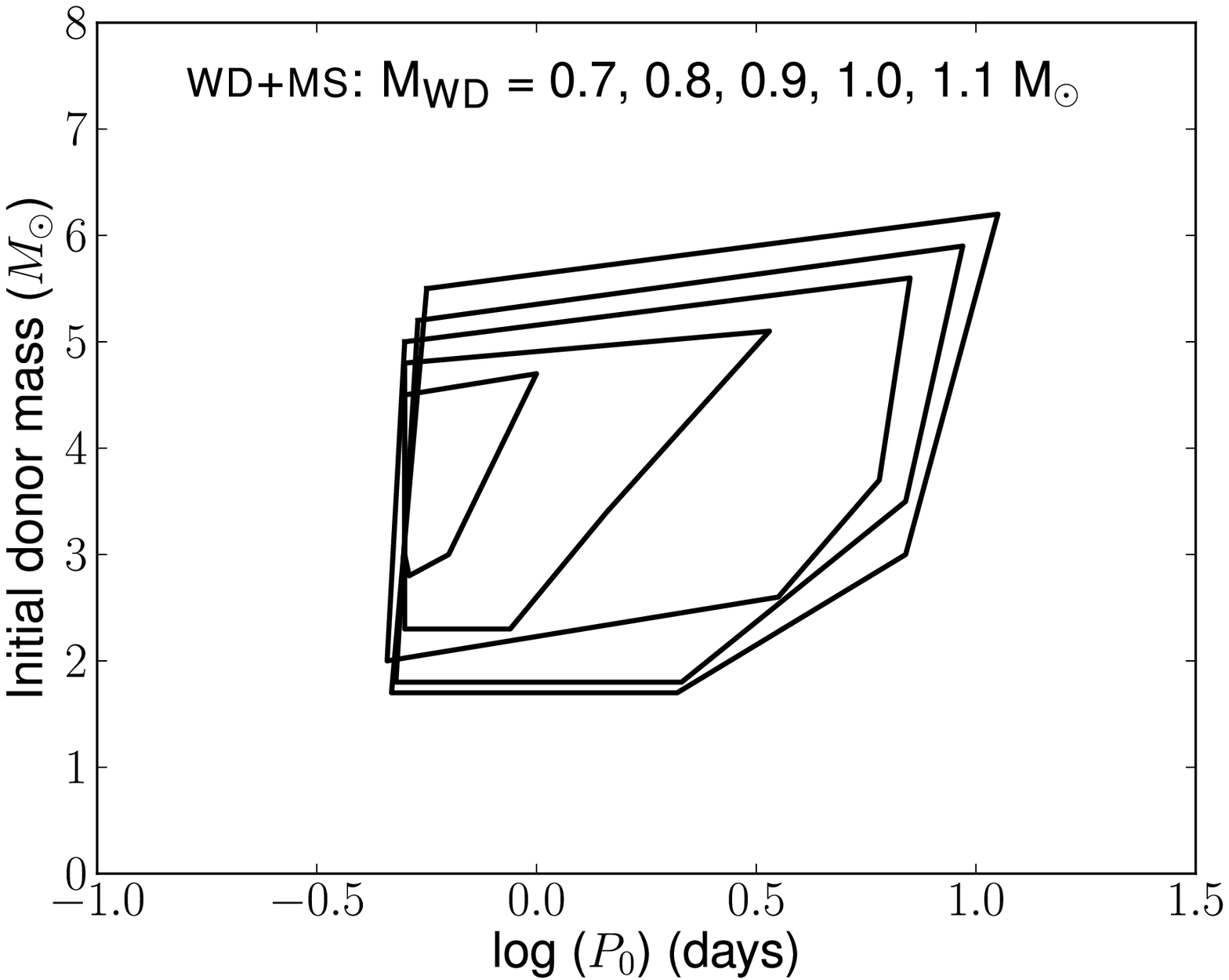}
\caption{Initial parameter regions for the WD+MS track to SNeIa. The different contours are for different WD masses (increasing in size with increasing mass) and the stripping parameter $c_1 = 3$. Left: from \citetalias{HKN08}. The hatched regions indicate SNIa explosions with short delay times of $t < 100$Myr for M = 0.7\Msolar~and M = 1.1\Msolar. Right: initial parameter regions as used in this work.}
\label{fig_ch2:WD+MS}
\end{figure}

\begin{figure}[!hbtp]
\centering
\includegraphics[width=0.24\textwidth]{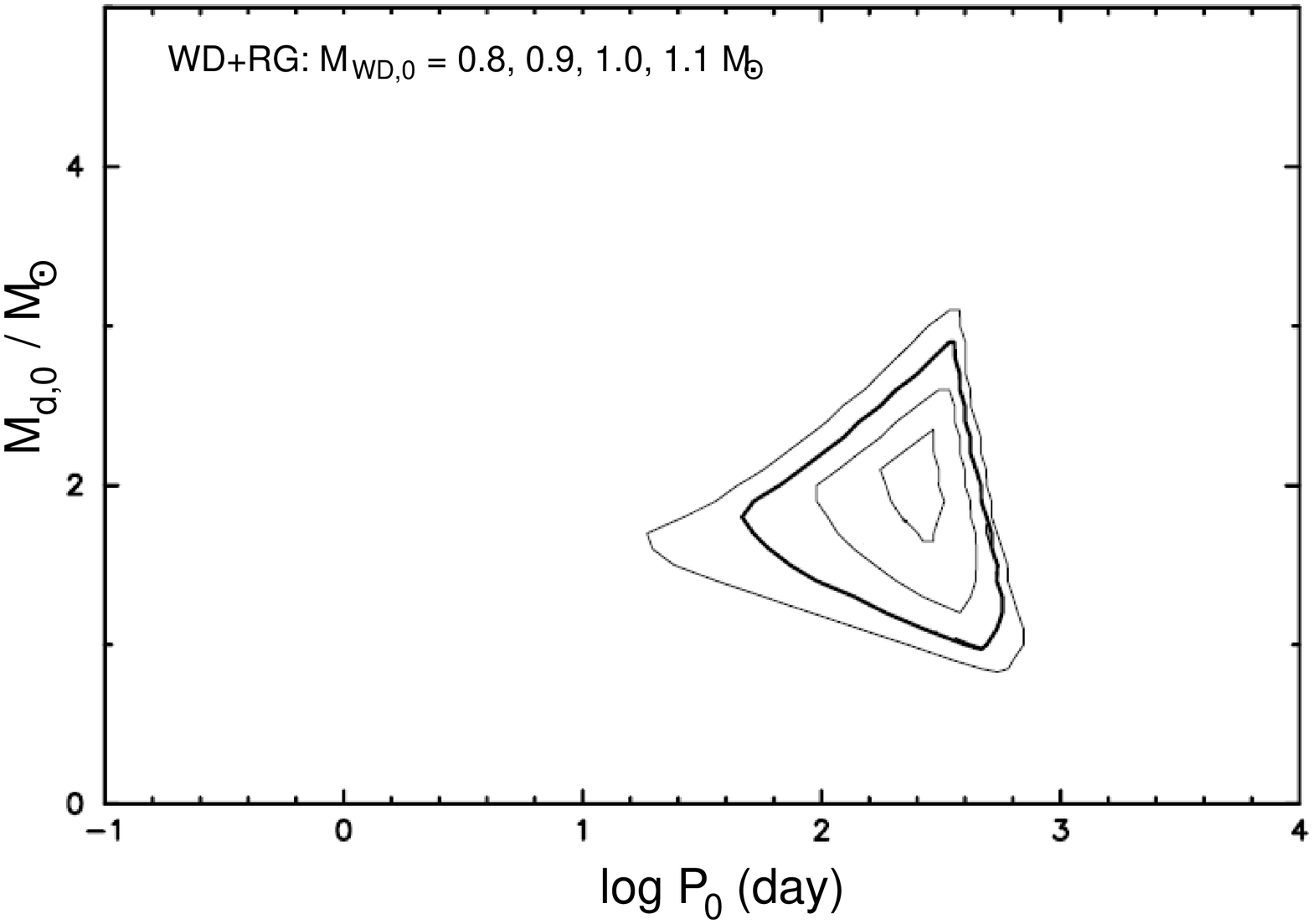}
\includegraphics[width=0.24\textwidth]{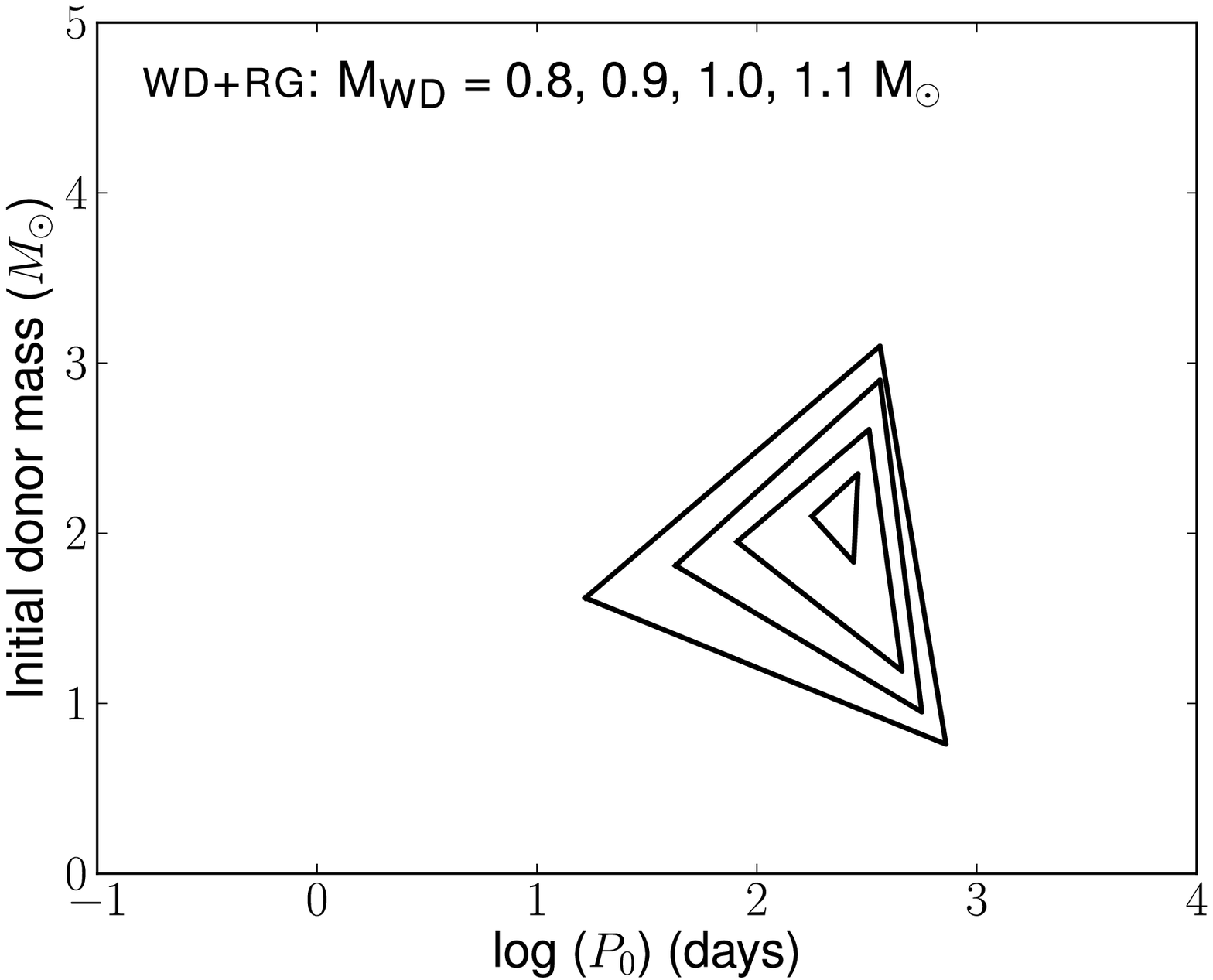}
\caption{Initial parameter regions for the WD+RG track to SNeIa. The different contours are for different WD masses (increasing in size with increasing mass) and the stripping parameter $c_1 = 1$. Left: from \citetalias{HKN99}. Right: initial parameter regions as used in this work.}
\label{fig_ch2:WD+RG}
\end{figure}

In this method, the progenitor binaries are selected on $P_\mathrm{orb}$, $M_\mathrm{WD}$, and $M_\mathrm{comp}$ at the time of WD birth (third stage in Fig.\,\ref{fig_ch2:track}). 
The parameter regions used are shown at the right hand side in Fig.\,\ref{fig_ch2:WD+MS} for binaries containing a WD and a MS star or Hertzsprung gap star (WD+MS binaries) and Fig.\,\ref{fig_ch2:WD+RG} for those containing a WD and a red giant star (WD+RG binaries). They are made to match the islands in \citetalias{HKN08} and \citetalias{HKN99} respectively, which are shown on the left hand side in these figures. For the WD+MS channels a moderate wind-stripping parameter of $c_1=3$ is taken (see eq.\,\ref{eq_ch2:stripping}). 

\section{Results: SNIa rates and delay times} 
\label{sec_ch2:results}
\begin{figure}[]
\resizebox{\hsize}{!}{\includegraphics{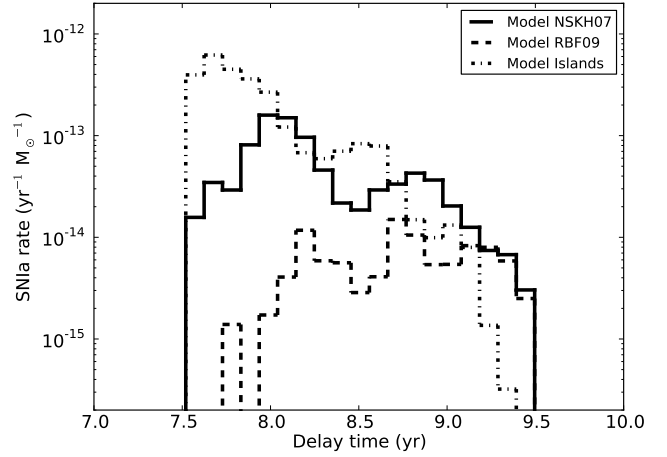}}
\caption{Delay time distribution assuming the $\gamma$-algorithm for the CE with $\gamma = 1.75$. Different lines correspond to the delay time distributions that result from four different approaches described in Sect\,\ref{sec_ch2:methods}, all for the $\gamma$-prescription. The dot-dashed line shows the result of the Islands selection method, the three other linestyles correspond to the retention efficiencies as in Fig.\,\ref{fig_ch2:reteffs_tot}. Note that for the retention efficiency of \citetalias{Y10} no SD binaries evolve into a SNeIa.}
\label{fig_ch2:SDDTDs}
\end{figure}

\begin{figure}[]
\resizebox{\hsize}{!}{\includegraphics{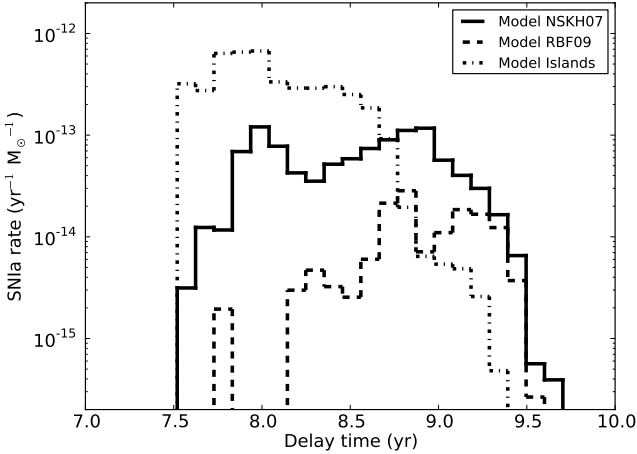}}
\caption{Same as in Fig.\,\ref{fig_ch2:SDDTDs} but assuming the $\alpha$-formalism with $\alpha\lambda = 0.5$. Again no SD binaries evolve into a SNeIa for the retention efficiency of \citetalias{Y10}.}
\label{fig_ch2:SDDTDs_alpha}
\end{figure}

We have simulated a population of SD SNIa progenitors for the four approaches described in Sect.\,\ref{sec_ch2:methods} assuming our model for the $\gamma$-prescription or the $\alpha$-prescription. The delay time distributions for the $\gamma$-prescription simulations can be seen in Fig.\,\ref{fig_ch2:SDDTDs} and for the $\alpha$-prescription in Fig.\,\ref{fig_ch2:SDDTDs_alpha}. The delay time $t$ is the time at which a SNIa occurs, where $t=0$ is the time when the binary is born as a double ZAMS-star binary. A DTD shows the distribution of delay times after a single starburst involving a large number of binaries. 
For model Islands, we assume the phase between the onset of mass transfer and the SNIa explosion is short compared to the lifetime of the binary. 

The progenitor islands are based on the work of \citetalias{NSKH07} and \citetalias{HKN08}, as is our model NSKH07 that uses the retention efficiencies directly. If we compare the DTD resulting from model Islands with model NSKH07, there is a noticeable difference at early delay times. The DTD of model Islands peaks earlier and higher. The reason is the extend of the Islands to high donor masses of about 6\Msolar for WD+MS progenitors. These massive companions fill their Roche lobes soon and therefore the SNIa explosion occurs at earlier delay times ($t < 1$Gyr). In SeBa, binaries with such high mass ratios undergo unstable mass transfer and a CE phase. They do not develop a SNIa explosion, resulting in less SNeIa at early delay times from the retention efficiencies method.

Our progenitor regions in model Island consist of adjacent slices that increase in size for increasing M$_\mathrm{WD}$, for both WD+MS and WD+RG progenitors. These slices have a thickness of 0.1\Msolar~around the values M$_\mathrm{WD}$ = 0.7, 0.8, 0.9, 1.0 and 1.1\Msolar~for WD+MS binaries and M$_\mathrm{WD}$ = 0.8, 0.9, 1.0 and 1.1\Msolar~for WD+RG binaries. Ideally the volume enclosing the progenitor binaries has smooth edges in all three dimensions. Since this is not the case in our approach we explored the resulting inaccuracy by increasing our resolution. We double the amount of islands covering the same range of WD mass, so the thickness of each island slice is now 0.05\Msolar. We found that the change in the final integrated SNIa rate is 3-4\% and the DTD is barely affected.

The SNIa DTD and integrated rate are strongly influenced by the prescription of the retention efficiency, as shown in Fig.\,\ref{fig_ch2:SDDTDs} and Fig.\,\ref{fig_ch2:SDDTDs_alpha}, and Table\,\ref{tbl_ch2:rates}. For more pessimistic values of the total retention efficiency the rates and overall height of the DTD decreases. Note that the total retention efficiency of model Y10 is so small that no SNeIa developed and only an upper limit is given in Table\,\ref{tbl_ch2:rates}. Model NSKH07 and RBF09 give rise to DTDs that differ most strongly at short delay times. This is due to high donor masses that transfer mass at a high rates, where the total retention efficiencies differ significantly, see Fig.\,\ref{fig_ch2:reteffs_tot}. This results from the inclusion of the wind stripping effect in \citetalias{NSKH07}.

Changing the prescription used for the CE phase modifies the integrated SNIa rate by a factor of about two. 
When the $\gamma$-prescription is applied, less (very) close binaries are created as the binary orbits do not shrink as effectively as for the $\alpha$-prescription. An exception to this is when the mass ratio of the binary is extreme at the moment of Roche lobe overflow. 
These systems merge at short delay times t$<$0.1Gyr. Therefore the DTD when the $\gamma$-prescription is assumed declines faster with time than when the $\alpha$-prescription is assumed. The classical evolution path towards a SD binary as depicted in Fig.\,\ref{fig_ch2:track}, is less common in the BPS model using the $\gamma$-prescription. Assuming the $\alpha$-prescription 80\% of the SD binaries evolve through this channel, where as for the  $\gamma$-prescription this has decreased to 30-40\% (depending on the retention efficiency). In the most common evolution path for the model assuming the $\gamma$-prescription, the first phase of mass transfer is stable. 
 
\begin{table*}[t]
\caption{Time-integrated SNIa rates in the SD channel for different combinations of the retention efficiency and the common envelope prescription in units of $10^{-4}$\Msolar$^{-1}$.} 
\centering
\begin{tabular}{l l c c}
\hline\hline
Model & Approach used & $\gamma$-prescription & $\alpha$-prescription \\
\hline
Model NSKH07 & Retention efficiency of \citetalias{NSKH07}$^1$ & 0.59 & 1.3 \\
Model RBF09 & Retention efficiency of \citetalias{Rea09}$^2$ & 0.19 & 0.35 \\
Model Y10 & Retention efficiency of \citetalias{Y10}$^3$ & $<$0.001 & $<$0.001 \\
Model Islands & Islands of \citetalias{HKN08}$^4$, \citetalias{HKN99}$^5$ & 0.73 & 1.5 \\   
\hline
\end{tabular}
\begin{flushleft}
\tablefoot{$^{1}$ \citet{NSKH07}; $^{2}$ \citet{Rea09}; $^{3}$ \citet{Y10}; $^{4}$ \citet{HKN08}; $^{5}$ \citet{HKN99}}
\end{flushleft}
\label{tbl_ch2:rates}
\end{table*}

\section{Discussion}
\label{discussion}

The theoretical SD SNIa rates that follow from SeBa in this study can be compared to the results of various other BPS research groups \citep[e.g. \citetalias{Rea09}; \citetalias{Y10};][]{Mea10, Wan10, Cea11}. Table\,\ref{tbl_ch2:snrates} shows the time-integrated SNIa rates of these groups taken from \citet{Nel12}. The disagreement in the rates is large up to a factor of about 600, but so far no explanation has been found. 
Note that \citet{Nel12} rescaled the results (if needed) to the same initial distribution of parameters as discussed in Sect.\,\ref{sec_ch2:methods}. Assumptions and simplifications vary between the different BPS codes \citetext{See Toonen et al. in prep. for a study on this} causing differences in their predictions. An important assumption for the BPS simulations is the assumed CE-prescription, which is also given in in Table\,\ref{tbl_ch2:snrates}. The effect of different values for $\alpha$ has been studied by several groups, e.g. the effect on the DTD \citep{Wan10, Rea09, Mea10, Cea11}. The effect on the integrated rate of a small change in $\alpha$ is of the order of a factor 0.7-3 \citep[][]{Rea09, Mea10}, however the effect can be up to an order of magnitude for larger changes in $\alpha$ (Claeys et al. in prep.).

The entries in Table\,\ref{tbl_ch2:snrates} are ordered to increase in rate. The smallest rate is from the work of Yungelson \citepalias[]{Y10}. Similarly we find the lowest rate for model Y10, however we find an even lower rate than Yungelson.  
The preliminary rate of \citet{Cea11} is significantly lower than our corresponding model. However \citet{Cea11} use a variable lambda prescription, whereas in our best corresponding model we set $\alpha\lambda=0.5$.
The integrated rates of \citet{Rea09} is a factor of about two lower than the rates of our best corresponding model. The rates of \citet{Mea10} and \citet{Wan10} are a factor of a about seven and two higher respectively. So although our results do recover the trend in the theoretical predictions from different binary population synthesis codes, they do not fully explain the large disagreement among them. 

The integrated rates based on observations are given in Table\,\ref{tbl_ch2:snrates}. The most recent measurements are based on field galaxies and generally show lower rates, while earlier estimates based on galaxy clusters are higher. At this moment it is unclear if the different observed integrated rates are due to systematic effects or if there is a real enhancement of SNeIa in cluster galaxies. See also \citet{Mao12} for a discussion. Even though the different retention efficiency models affect the SNIa rates with a factor $>10^3$, none of the integrated rates comfortably matches the observed rates, especially those from galaxy clusters. In addition, the DTD reconstructed from observations typically show a continuation to longer delay times, which are absent in all our SD DTDs. We conclude from this, that in the current model of SD SNIa theory, the main contribution to the SNIa rate comes from other evolution channels. One possible channel involves semi-detached binaries in which a WD accretes from a hydrogen-poor helium-rich donor, such as sdB stars. \citet{Wan09}, \citet{Rea09} and \citet{Cea11} showed that in this channel the DTD peaks at delay times of about 100Myr, although rates at this delay time vary between $10^{-4}-10^{-2} \mathrm{yr}^{-1} (10^{10}$\Msolar$)^{-1}$. The contribution from binaries of the DD channel is debated heavily. Explosion models favour accretion-induced collapse to a neutron star over a SNIa event \citep[see e.g.][for a review]{Hil00}. However, BPS codes find more SNIa events from the standard DD channel than the SD channel \citep[\citetalias{Rea09}; \citetalias{Y10};][]{Mea10, Too12}\footnote{Note that the integrated rate from the violent merger model for double WDs \citep{Pak10,Pak11,Pak12,Rop12} can be much lower \citep{Too12, Che12, Rui12}}. \citet{Too12} study the contribution from the double-degenerate channel with SeBa comparing the $\alpha$- and $\gamma$-prescription for the CE. They find that even though the DD DTDs fit the observed DTD beautifully, the normalisation does not by a factor of about 7-12 compared to the cluster rates. Taking into account the new rates from field galaxies, the factor becomes about 1.2-12. Other contributions to the SNIa rate can possibly come from e.g. core-degenerate mergers \citep{Kas11}, double detonating sub-Chandrasekhar accretors \citep[see e.g.][]{Kro10} or Kozai oscillations in triple systems \citep{Sha12, Ham13}.

\begin{table*}[]
\caption{Time-integrated SNIa rates in the SD channel for the assumptions of different research groups in units of $10^{-4}$\Msolar$^{-1}$}
\label{tbl_ch2:snrates}
\centering
\begin{tabular}{c c c c |c c}
\hline\hline
BPS & WD accretion & SNIa rate  & CE &  SNIa rate &  CE                             \\
  \hline
Yungelson$^{1}$ & \citetalias{Y10}                             & $<$0.001 & $\gamma=1.5 $       & 0.006     & $\gamma=1.5$                    \\
Claeys$^{2}$    & \citetalias{HKN99}                           & 1.3      & $\alpha\lambda=0.5$ & 0.13      & $\alpha=1$, $\lambda=variable$  \\
Ruiter$^{3}$    & \citetalias{Rea09}                           & 0.35     & $\alpha\lambda=0.5$ & 0.17      & $\alpha\lambda=0.5$             \\
Wang/Han$^{4}$  & \citetalias{HKN99}                           & 1.3      & $\alpha\lambda=0.5$ & 2.8       & $\alpha\lambda = 0.5$           \\
Mennekens$^{5}$ & \citetalias{HKN99}, \citetalias{HKN08}       & 0.55     & $\alpha\lambda=1$   & 3.7       & $\alpha\lambda = 1$             \\
\hline
Observed$^{6}$  &  & 4-26 & & & \\
\hline
\end{tabular}
\begin{flushleft}
\tablefoot{Columns~3~and~5 shows the integrated rate predicted by SeBa and the BPS group in question respectively. The assumption of the BPS group for the CE-evolution is shown in column~6 and the best corresponding CE-model in SeBa is shown in column~4. $^{1}$ \citetalias{Y10}; $^{2}$ \citet{Cea11} (preliminary results); $^{3}$ \citetalias{Rea09}; $^{4}$ \citet{Wan10}; $^{5}$ \citet{Mea10}; $^{6}$ \citet{Mao11b, Per12, Mao12, Gra12}}
\end{flushleft}
\end{table*}

\section{Conclusions} 
\label{sec_ch2:conclusions}
In this work we have studied the effect of the poorly understood phase of WD accretion in the context of supernova type Ia rates. We employed the binary population synthesis code SeBa \citep{PZV96, Nea01, Too12} to study the SNIa rates and progenitors for different CE prescriptions. We differentiated between four models assuming either one of three retention efficiencies of \citetalias{NSKH07}, \citetalias{Rea09} and \citetalias{Y10} or making a selection of SNIa progenitors based on binary parameters at the time of WD formation \citepalias{HKN08, HKN99}. The three retention efficiencies assumed by different binary population synthesis codes differ strongly. The difference comes from a lack of understanding at low mass transfer rates where novae occur, and mass transfer rates higher than the rate for stable burning. This is true for the accretion of hydrogen that is transferred to the WD by the companion, as well as the accretion of helium that has been burned from hydrogen on the WD. The efficiency with which a C/O WD grows in mass is strongly affected by the combination of the efficiency of hydrogen and helium accretion. For example the hydrogen and helium retention efficiencies of \citetalias{Y10} are maximal at different ranges of the mass transfer rate resulting in a low total retention efficiency. 

The total number of SNIa progenitors is significantly influenced by the choice of the model. The integrated SNIa rate vary between $<1\cdot10^{-7}-1.5\cdot10^{-4}$\Msolar$^{-1}$, where the rates are highest for the model that assume the retention efficiencies of \citetalias{NSKH07} and only an upper limit can be given for \citetalias{Y10}. Our method based on the parameter space of binaries at WD birth of \citetalias{HKN08} and \citetalias{HKN99} consists of a discrete set of islands in ($P_\mathrm{orb}$ - $M_\mathrm{WD}$ - $M_\mathrm{comp}$). The discretisation introduces an error of 3-4\% on the integrated rates. We showed in this paper that independent of the WD accretion model the SNIa rate approximately doubles when the $\alpha$-prescription is assumed with $\alpha\lambda = 0.5$ as compared to the $\gamma$-formalism with $\gamma=1.75$. The effect of different values for $\alpha$ on the SD SNIa rate is a factor of 0.7-3 \citep{Rea09, Mea10} for small changes in the CE-efficiency and up to an order of magnitude for larger changes in the CE-efficiency (Claeys et al. in prep.). Also note that throughout this work we have assumed solar values for the stellar metallicities. Using a broad range of metallicities in the BPS code might influence the SNIa rates and DTDs, both in the Islands and retention efficiency approach. See \citet{KTNHK98} for a study on how the progenitor islands depend on metallicity.

Several predictions for the SD rate have been made by different groups using binary population synthesis simulations. The results show a wide spread and do not agree with each other or with observational data. The integrated rates vary between $6\cdot10^{-7}-3.7\cdot10^{-4}$\Msolar$^{-1}$. In this study we find that also the model for the WD accretion is a major source of uncertainty on the SNIa rates. The different prescriptions for the retention efficiency introduce a source of uncertainty with an effect on the integrated rates by a factor of about 3-4 (comparing our models that assume the retention efficiencies of \citetalias{NSKH07} with \citetalias{Rea09}) or even larger up to a factor of a few hundred (comparing with the efficiency of \citetalias{Y10}). Although our results do recover the trend in the theoretical predictions from different binary population synthesis codes, they do not fully explain the large disagreement among them. As the exact origin of the differences in the SD rate remains unclear, Toonen et al. in prep. study the results of four different binary population synthesis codes and investigate the importance of the different assumptions and numerical approaches in these codes.

\begin{acknowledgements}
We thank Zhanwen Han for useful comments which helped us to improve the manuscript. This work was supported by the Netherlands Research Council NWO (grant VIDI [\# 639.042.813]) and by the Netherlands Research School for Astronomy (NOVA).
\end{acknowledgements}

\bibliographystyle{aa}
\bibliography{bibliography.bib}

\appendix
\section{Retention efficiencies} 
\label{app_ch2:retentionefficiencies}
The total retention efficiency is the product of the hydrogen and helium retention efficiencies. In the equations in this appendix, all $\dot{M}$ are in units of \Msolar/yr. 

\subsection{Retention efficiencies based on \citetalias{NSKH07}}
The hydrogen retention efficiency is a strong function of the mass transfer rate of the companion $\dot M_\mathrm{comp}$. Three regimes can be distinguished, separated by the stable and critical mass transfer rates ($\dot{M}_\mathrm{st}$ and $\dot{M}_\mathrm{cr}$), see Table\,\ref{tbl_ch2:regimes_nomoto}. In the nova and stable burning regime all the mass lost by the companion is transferred onto the WD. In the stable regime all hydrogen-rich matter is burned into helium-rich matter and all stays on the WD, so that $\eta_{\mathrm{H}}=1$. The nova regime is linearly interpolated between the lower boundary at $|\dot M_\mathrm{comp}| = 10^{-7}$ and the start of the stable regime at $\dot{M}_\mathrm{st}$. In the third regime the nuclear hydrogen burning on top of the white dwarf is so strong that a wind is produced which not only attenuates the mass transfer rate but can also strip the companion's outer envelope. It is no longer the case that all the mass lost by the companion accretes onto the white dwarf. The maximum that the white dwarf can accrete is $\dot{M}_\mathrm{cr}$, which is a fraction $\eta_\mathrm{H}$ of the mass transferred to the white dwarf $\dot{M}_\mathrm{H}$. All the excess is blown off by the wind. The amount of matter that is stripped from the companion is defined by eq.\,\ref{eq_ch2:stripping} and the stripping parameter is taken to be $c_1=3$.

\begin{table}[!htb]
\label{tbl_ch2:regimes_nomoto}
\caption{The three regimes for different mass transfer rates of the companion star, as in \citetalias{NSKH07}.}
\centering
\begin{tabular}{l c}
\hline\hline
\rule{0pt}{2.5ex}  & $\dot M_\mathrm{comp}$ range \\
\hline
\rule{0pt}{2.5ex} Nova regime & $|\dot{M}_\mathrm{comp}| < \dot{M}_\mathrm{st}$ \\
\rule{0pt}{2.5ex} Stable burning regime & $\dot{M}_\mathrm{st} \leq |\dot{M}_\mathrm{comp}| \leq \dot{M}_\mathrm{cr}$ \\
\rule{0pt}{2.5ex} Wind and stripping regime & $|\dot{M}_\mathrm{comp}| > \dot{M}_\mathrm{cr}$ \\
\hline
\end{tabular}
\end{table}

The exact prescriptions for the stable and critical mass transfer rates are:
\begin{equation}
\dot{M}_\mathrm{st} = 3.1 \cdot 10^{-7} \Big(\frac{M_\mathrm{WD}}{\mathrm{M}_{\odot}}-0.54\Big) \mathrm{~ and}
\end{equation}
\begin{equation}
\dot{M}_\mathrm{cr} = 7.5 \cdot 10^{-7} \Big(\frac{M_\mathrm{WD}}{\mathrm{M}_{\odot}}-0.40\Big).
\end{equation}

For the hydrogen retention efficiency we arrive at the following:
\begin{equation}
\eta_\mathrm{H} = \left\{ 
\begin{array}{rl}
0 &\mathrm{~ if ~ ~} |\dot{M}_\mathrm{comp}| < 10^{-7} \\
(\mathrm{log}(|\dot{M}_\mathrm{comp}|)~+&7)/((\mathrm{log}(\dot{M}_\mathrm{st}) + 7 ) \\
 &\mathrm{~ if ~ ~} 10^{-7} < |\dot{M}_\mathrm{comp}| < \dot{M}_\mathrm{st} \\
1 &\mathrm{~ if ~ ~} \dot{M}_\mathrm{st} < |\dot{M}_\mathrm{comp}| < \dot{M}_\mathrm{cr} \\
\dot{M}_\mathrm{cr}/\dot{M}_\mathrm{H} &\mathrm{~ if ~ ~} \dot{M}_\mathrm{cr} < |\dot{M}_\mathrm{comp}| < 10^{-4}
\label{eq_ch2:NomotoRetEffH}
\end{array} \right.
\end{equation}

with, for the wind and stripping regime,
\begin{equation}
\frac{\dot{M}_\mathrm{cr}}{\dot{M}_\mathrm{H}} = \frac{(c_1+1)\dot{M}_\mathrm{cr}}{c_1\dot{M}_\mathrm{cr}+|\dot{M}_\mathrm{comp}|}.
\end{equation}

This last equation follows from some algebra, taking into account that $\dot{M}_\mathrm{comp}, \dot{M}_\mathrm{wind} \mathrm{~ and ~} \dot{M}_\mathrm{strip} < 0$ because they describe matter travelling away from one of the stars and using eq.\,\ref{eq_ch2:stripping}. 

For the helium retention efficiency the following prescriptions were used:
\begin{equation}
\eta_\mathrm{He} = \left\{ 
\begin{array}{rl}
0 &\mathrm{~ if ~ ~} \dot{M}_\mathrm{He} < 10^{-7.8} \\
-0.175(\mathrm{log}(\dot{M}_\mathrm{He})~+&5.35)^2+1.05 \\ &\mathrm{~ if ~ ~} 10^{-7.8} < \dot{M}_\mathrm{He} < 10^{-5.9} \\
1 &\mathrm{~ if ~ ~} 10^{-5.9} < \dot{M}_\mathrm{He} < 10^{-5.0} \\
0 &\mathrm{~ if ~ ~} \dot{M}_\mathrm{He} > 10^{-5.0} 
\label{eq_ch2:NomotoRetEffHe}
\end{array} \right.
\end{equation}
where $\dot{M}_\mathrm{He} = \eta_\mathrm{H} \dot{M}_\mathrm{comp}$.

\subsection{Retention efficiencies based on \citetalias{Rea09}.} 
\label{app_ch2:reteff_ruiter}
For the hydrogen retention efficiency two regimes can be distinguished, the nova regime and the wind regime, see Table\,\ref{tbl_ch2:regimes_ruiter}. The stripping effect is not taken into account ($c_1=0$). In the hydrogen nova regime an interpolation of the results from \citet{PK95} is used for $\eta_{\mathrm{H}}$. The amount of mass transferred to the white dwarf is always equal to the amount of mass lost by the companion, $|\dot{M}_\mathrm{H}| = |\dot{M}_\mathrm{comp}|$. 

\begin{table}[!htb]
\label{tbl_ch2:regimes_ruiter}
\caption{The two regimes for different mass transfer rates of the companion star, as in \citetalias{Rea09}.}
\centering
\begin{tabular}{l c}
\hline\hline
\rule{0pt}{2.5ex}  & $\dot M_\mathrm{comp}$ range \\
\hline
\rule{0pt}{2.5ex} Nova regime & $|\dot{M}_\mathrm{comp}| < \dot{M}_\mathrm{cr}$ \\
\rule{0pt}{2.5ex} Wind regime & $|\dot{M}_\mathrm{comp}| > \dot{M}_\mathrm{cr}$ \\
\hline
\end{tabular}
\end{table}

The critical mass transfer rate is given by: 
\begin{equation}
\dot{M}_\mathrm{cr} = 7.5 \cdot 10^{-7} \Big(\frac{M_\mathrm{WD}}{\mathrm{M}_{\odot}}-0.40\Big).
\label{eq_ch2:RuiterMdot}
\end{equation}

The hydrogen retention efficiency is:
\begin{equation}
\eta_\mathrm{H} = \left\{ 
\begin{array}{rl}
0 &\mathrm{~ if ~ ~} |\dot{M}_\mathrm{comp}| < 10^{-8} \\
\mathrm{0.25(log(|\dot{M}_\mathrm{comp}|)+8) } &\mathrm{~ if ~ ~} 10^{-8} < |\dot{M}_\mathrm{comp}| < 10^{-7} \\
0.25 + 0.75(\mathrm{log}(|\dot{M}_\mathrm{comp}|)~+&7)/(\mathrm{log}(\dot{M}_\mathrm{cr}) +7 )\\
&\mathrm{~ if ~ ~} 10^{-7} < |\dot{M}_\mathrm{comp}| < \dot{M}_\mathrm{cr} \\
\dot{M}_\mathrm{cr}/\dot{M}_\mathrm{H} &\mathrm{~ if ~ ~} \dot{M}_\mathrm{cr} < |\dot{M}_\mathrm{comp}| < 10^{-4}
\label{eq_ch2:RuiterRetEffH}
\end{array} \right.
\end{equation}

where 
\begin{equation}
\frac{\dot{M}_\mathrm{cr}}{\dot{M}_\mathrm{H}} = \frac{\dot{M}_\mathrm{cr}}{|\dot{M}_\mathrm{comp}|}.
\end{equation}

The helium retention efficiency is given by 
\begin{equation}
\eta_\mathrm{He} = \left\{ 
\begin{array}{rl}
0 &\mathrm{~ if ~ ~} \dot{M}_\mathrm{He} < 10^{-7.3} \\
-0.175(\mathrm{log}(\dot{M}_\mathrm{He})~+&5.35)^2+1.05 \\ 
&\mathrm{~ if ~ ~} 10^{-7.3} < \dot{M}_\mathrm{He} < 10^{-5.9} \\
1 &\mathrm{~ if ~ ~} 10^{-5.9} < \dot{M}_\mathrm{He} < 10^{-5.0} \\
0 &\mathrm{~ if ~ ~} \dot{M}_\mathrm{He} > 10^{-5.0} 
\label{eq_ch2:RuiterRetEffHe}
\end{array} \right.
\end{equation}
where $\dot{M}_\mathrm{He} = \eta_\mathrm{H} \dot{M}_\mathrm{comp}$. Note the similarity with the helium retention efficiency of \citetalias{NSKH07}, except for the lower limit of the mass transfer rate of the stable burning regime.

\subsection{Retention efficiencies based on \citetalias{Y10}.}
Two regimes can be distinguished for the hydrogen retention efficiency, the nova and wind regimes, see Table\,\ref{tbl_ch2:regimes_yungelson}. The same interpolation from \citet{PK95} is used for the hydrogen nova regime. The stripping effect is not taken into account ($c_1=0$). In all cases the amount of mass transferred to the white dwarf is equal to the amount of mass lost by the companion, $|\dot{M}_\mathrm{H}| = |\dot{M}_\mathrm{comp}|$. It is very similar to the retention efficiencies in Sect.\,\ref{app_ch2:reteff_ruiter}, except that the prescription for $\dot{M}_\mathrm{cr}$ is different.

\begin{table}[!htb]
\label{tbl_ch2:regimes_yungelson}
\caption{The two regimes for different mass transfer rates of the companion star, as in \citetalias{Y10}.}
\centering
\begin{tabular}{l c}
\hline\hline
\rule{0pt}{2.5ex}  & $\dot M_\mathrm{comp}$ range \\
\hline
\rule{0pt}{2.5ex} Nova regime & $|\dot{M}_\mathrm{comp}| < \dot{M}_\mathrm{cr}$ \\
\rule{0pt}{2.5ex} Wind regime & $|\dot{M}_\mathrm{comp}| > \dot{M}_\mathrm{cr}$ \\
\hline
\end{tabular}
\end{table}

The prescription for the critical mass transfer rate is:
\begin{equation}
\dot{M}_\mathrm{cr} = 10^{-9.31+4.12M_\mathrm{WD}-1.42M_\mathrm{WD}^2}.
\label{eq_ch2:YungelsonMdot}
\end{equation}

The hydrogen retention efficiency is given by:
\begin{equation}
\eta_\mathrm{H} = \left\{ 
\begin{array}{rl}
0 &\mathrm{~ if ~ ~} |\dot{M}_\mathrm{comp}| < 10^{-8} \\
\mathrm{0.25(log(|\dot{M}_\mathrm{comp}|)+8) } &\mathrm{~ if ~ ~} 10^{-8} < |\dot{M}_\mathrm{comp}| < 10^{-7} \\
\mathrm{0.25 + 0.75 (log(|\dot{M}_\mathrm{comp}|)} ~+&7)/ \mathrm{(log(\dot{M}_\mathrm{cr}) +7 ) } \\
 &\mathrm{~ if ~ ~} 10^{-7} < |\dot{M}_\mathrm{comp}| < \dot{M}_\mathrm{cr} \\
\dot{M}_\mathrm{cr}/\dot{M}_\mathrm{H} &\mathrm{~ if ~ ~} \dot{M}_\mathrm{cr} < |\dot{M}_\mathrm{comp}| < 10^{-4}
\label{eq_ch2:YungelsonRetEffH}
\end{array} \right.
\end{equation}
with
\begin{equation}
\frac{\dot{M}_\mathrm{cr}}{\dot{M}_\mathrm{H}} = \frac{\dot{M}_\mathrm{cr}}{|\dot{M}_\mathrm{comp}|}.
\end{equation}

The helium retention efficiency is:
\begin{equation}
\eta_\mathrm{He} = \left\{ 
\begin{array}{rl}
0 &\mathrm{~ if ~ ~} \dot{M}_\mathrm{He} < 10^{-7.5} \\
\frac{\dot{M}_\mathrm{He}}{10^{-5.75}} &\mathrm{~ if ~ ~} 10^{-7.5} < \dot{M}_\mathrm{He} < 10^{-5.7} \\
0.95 &\mathrm{~ if ~ ~} 10^{-5.7} < \dot{M}_\mathrm{He} < 10^{-5.4} \\
\frac{10^{-5.45}}{\dot{M}_\mathrm{He}} &\mathrm{~ if ~ ~} 10^{-5.4} < \dot{M}_\mathrm{He} < 10^{-4.0} \\
0 &\mathrm{~ if ~ ~} \dot{M}_\mathrm{He} > 10^{-4.0} 
\label{eq_ch2:YungelsonRetEffHe}
\end{array} \right.
\end{equation}
where $\dot{M}_\mathrm{He} = \eta_\mathrm{H} \dot{M}_\mathrm{comp}$.

\end{document}